\documentstyle[psfig,12pt]{article}
\newlength{\titlesep}
\newlength{\authorsep}
\makeatletter

\def\fnum@figure{FIG.~\thefigure}

\newcounter{figureparent}
\@addtoreset{figure}{figureparent}

\@addtoreset{equation}{section}

\newcounter{eqnparent}
\@addtoreset{equation}{eqnparent}

\renewcommand{\abstract}{\if@twocolumn
  \section*{Abstract}
  \else
  \begin{center}
    {\bf Abstract\vspace{-.5em}\vspace{0pt}}
  \end{center}
  \quotation
  \fi}
\renewcommand{\endabstract}{\if@twocolumn\else\endquotation\fi}

\newcommand{\thismonth}{\ifcase\month\or
 January\or February\or March\or April\or May\or June\or
 July\or August\or September\or October\or November\or December\fi
 \space \number\year}

\newcommand{\preprintnumber}[1]
{\begin{flushright}
  \begin{tabular}{l} #1 \end{tabular}
  \end{flushright}}

\makeatother

\newcommand{\Rn}[1]{{\uppercase\expandafter{\romannumeral#1}}}
\newcommand{\gsim}%
{\mathrel{\mbox{\raisebox{-1.0ex}%
{$\stackrel{\textstyle >}{\textstyle \sim}$}}}}
\newcommand{\lsim}%
{\mathrel{\mbox{\raisebox{-1.0ex}%
{$\stackrel{\textstyle <}{\textstyle \sim}$}}}}
\newcommand{\ovl}[1]{\overline{#1}}
\newcommand{\wt}[1]{\widetilde{#1}}

\newcommand{\Journal}[4]{{#1} {\bf #2}, {#4} {(#3)}}

\newcommand{\plb}{\sl Phys.~Lett.~{\bf B}}

\newcommand{\pr}{\sl Phys.~Rev.}
\newcommand{\prd}{\sl Phys.~Rev.~{\bf D}}
\newcommand{\prl}{\sl Phys.~Rev.~Lett.}

\newcommand{\npb}{\sl Nucl.~Phys.~{\bf B}}

\newcommand{\ibid}{\it ibid.}

\newcommand{\epsfile}[1]{\relax}

\begin{document}
\baselineskip 18pt
\begin{titlepage}
\preprintnumber{%
KEK-TH-535 \\
KEK Preprint 97-125\\
}
\vspace*{\titlesep}
\begin{center}
{\LARGE\bf
CP violation in the $\mu \rightarrow 3\,e$ process and supersymmetric
grand unified theory
}
\\
\vspace*{\titlesep}
{\large Yasuhiro  Okada}\footnote{okaday@theory.kek.jp},\\
{\large Ken-ichi Okumura}\footnote{okumura@theory.kek.jp}\\
and
{\large Yasuhiro  Shimizu}\footnote{shimizuy@theory.kek.jp}\\
\vspace*{\authorsep}
{\it Theory Group, KEK, Tsukuba, Ibaraki, 305 Japan }
\\
\end{center}
\vspace*{\titlesep}
\begin{abstract}
A triple vector correlation in the 
$\mu^{+} \rightarrow e^{+} e^{+} e^{-}$ decay with polarized muons is
investigated as a probe to CP violating coupling constants in
supersymmetric models. A sizable triple correlation can be induced due
to a complex phase in the supersymmetric soft-breaking terms in the
SU(5) grand unified theory. Correlation with the electric dipole moments 
of electron and neutron are investigated and it is shown that these
quantities give independent information on possible CP violating sources.  
\end{abstract}
\hspace{1cm}PACS numbers: 11.30.Fs, 11.30.Er, 11.35.Bv, 12.60.Jv
\end{titlepage}
Lepton flavor violating (LFV) processes such as 
$\mu^{+} \rightarrow e^{+} \gamma$, 
$\mu^{+} \rightarrow e^{+} e^{+} e^{-}$ 
and $\mu^-$-$e^-$ conversion in atoms have important implications in
search for physics beyond the Standard Model (SM). 
Many extensions of the SM predict measurable
rates for these LFV processes. In particular it has been pointed out
that rates for these processes can be as large as or just below the
present experimental upper bounds in supersymmetric (SUSY) grand unified 
theories (GUTs) \cite{BH}.
Current experimental bounds for these processes are
${\bf B}(\mu^+ \rightarrow e^+ \gamma) \leq 4.9 \times
10^{-11}$ \cite{Bol},
${\bf B}(\mu^+ \rightarrow e^+e^+e^-) \leq 1.0 \times
10^{-12}$ \cite{sin},
$\sigma(\mu^-\, {\rm Ti} \rightarrow e^- \,{\rm Ti})
/\sigma(\mu^-\,\rm{Ti} \rightarrow
\rm{capture}) \leq 4.3 \times 10^{-12}$ \cite{sin2}.
Further improvements of these bounds in two or three orders of magnitude
will have a great impact on search for unified theories based on SUSY.

In this letter, we consider the $\mu^{+} \rightarrow e^{+} e^{+} e^{-}$ 
process with polarized muons.
If the initial muon is polarized we can define T-odd triple vector
correlation, $\vec{\sigma}\cdot (\vec{p}_1\times\vec{p}_2)$ where 
$\vec{\sigma}$ is the muon spin and $\vec{p}_1$, $\vec{p}_2$ are 
two independent momenta of the final particles. In this way we can 
measure CP and T nonconserving effects in LFV interactions \cite{m3eCP}. 
As possible sources of CP violation we consider SUSY soft-breaking 
terms in SUSY GUT. We show that it is possible to
generate a measurable T-odd asymmetry in this model.
We also consider the electron and the neutron electric dipole moments (EDMs) 
which are sensitive to the same CP violating phase and show these 
observables give independent information on the possible CP violating
sources.

Let us begin with the following effective lagrangian relevant for the
$\mu^{+} \rightarrow e^{+} e^{+} e^{-}$ decay.
\begin{eqnarray}
  \label{lagrangian}
  {\cal L} 
&=& m_\mu A_R \,\ovl{\mu_R} \sigma^{\mu\nu} e_L \, F_{\mu\nu}
+ m_\mu A_L \,\ovl{\mu_L} \sigma^{\mu\nu} e_R \, F_{\mu\nu}
\nonumber \\
&+& g_1 \,\ovl{\mu_R} e_L \, \ovl{e_R} e_L
+ g_2\, \ovl{\mu_L} e_R \, \ovl{e_L} e_R
+ g_3\, \ovl{\mu_R} \gamma^\mu e_R \, \ovl{e_R} \gamma_\mu e_R
\nonumber\\
&+& g_4\, \ovl{\mu_L} \gamma^\mu e_L \, \ovl{e_L} \gamma_\mu e_L
+ g_5\, \ovl{\mu_R} \gamma^\mu e_R \, \ovl{e_L} \gamma_\mu e_L
+ g_6\, \ovl{\mu_L} \gamma^\mu e_L \, \ovl{e_R} \gamma_\mu e_R
+ {\rm H.C}.
\end{eqnarray}
Here $A_R$, $A_L$ and $g_i$'s are in general complex coupling
constants. In order to obtain the above expression we have used the
Fiertz rearrangement for four-fermion terms and neglect terms
suppressed by the electron mass compared to the muon mass.

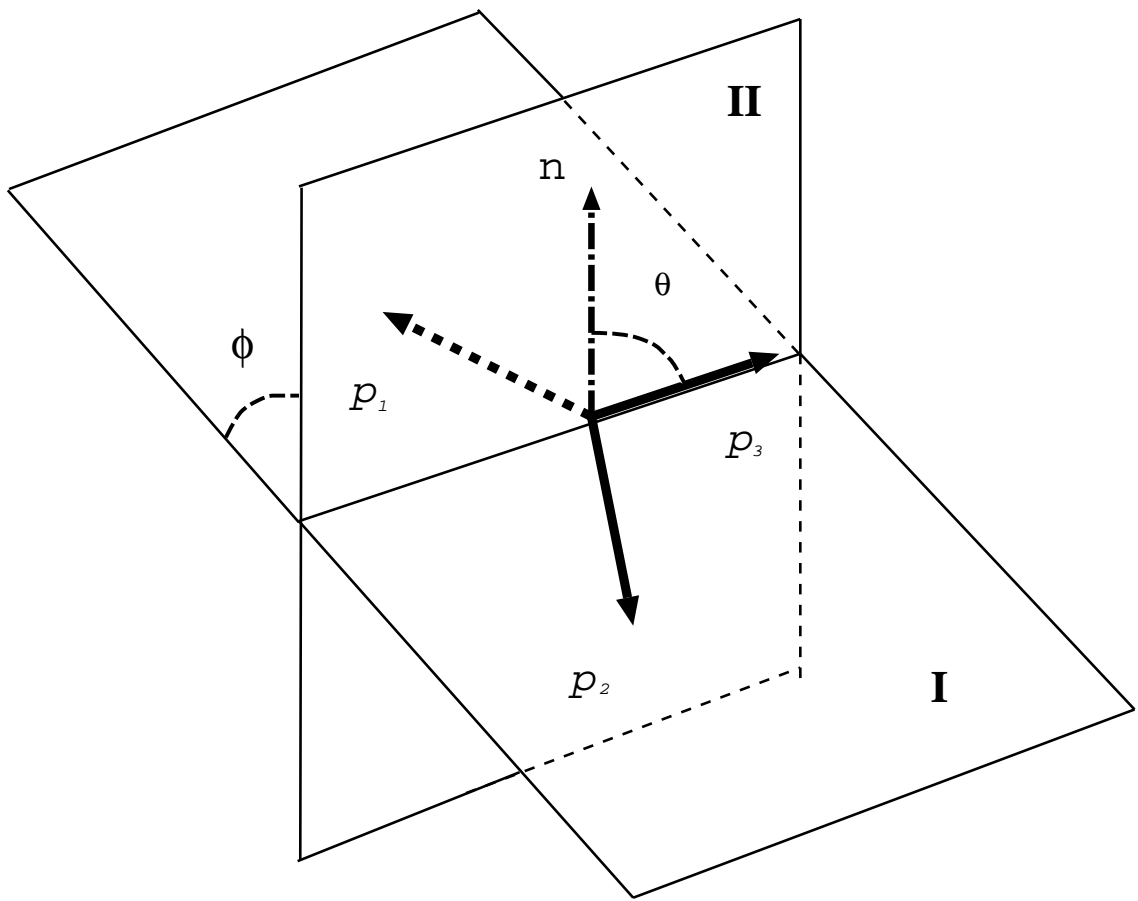
\begin{figure}[htbp]
  \begin{center}
    \leavevmode
    \caption{Kinematics of $\mu \rightarrow 3e$ decay in the center-of-mass
      system of muon. The plane I represents the decay plane on which 
      the momentum vectors $\vec{p}_1$, $\vec{p}_2$, $\vec{p}_3$ lie,
      where $\vec{p}_1$ and $\vec{p}_2$ are  momenta of two $e^+$'s
      and  $\vec{p}_3$ is momentum of $e^-$ respectively.
      The plane II is the plane which the muon polarization vector
      $\vec{n}$ and $\vec{p}_3$ make. }
    \label{fig:coordinate}
  \end{center}
\end{figure}

Among the above terms the photon-penguin terms,  $A_R$ and $A_L$,
induce the $\mu^{+} \rightarrow e^{+} \gamma$ decay in addition to
the $\mu^{+} \rightarrow e^{+} e^{+} e^{-}$ decay.
The branching ratio for $\mu^{+} \rightarrow e^{+} \gamma$ is given by
${\bf B}(\mu^{+} \rightarrow e^{+} \gamma) = \frac{48\pi^2}{G_F^2}\left(
\left| A_R \right|^2 + \left| A_L \right|^2 \right)$.
On the other hand the $\mu \rightarrow 3e$ branching ratio 
depends on four-fermion coupling constants $g_i$'s as well as $A_R$ 
and $A_L$. In order to present the differential branching ratio for
this process  we first discuss $\mu \rightarrow 3e$
kinematics with polarized muons which is specified by two energy valuables 
and two angle variables.
We assign  $p$, $p_1$, $p_2$ and $p_3$ to the momenta of $\mu^+$ and 
two $e^+$'s  and $e^-$, then two energy variables are taken to be
$x_1=\frac{2E_1}{m_\mu}$ and $x_2=\frac{2E_2}{m_\mu}$ . Two angle variables 
are necessary to specify the relative position between the muon 
polarization direction $(\vec{n})$ and the decay plane. Defining  $\vec{p}_1$
for the momentum of the positron which satisfies $(\vec{n} \times
\vec{p}_3 ) \cdot \vec{p}_1>0$, angle variables  $(\theta, \phi)$
can be identified as a polar coordinate of $\vec{n}$ 
($0 \leq \theta \leq \pi$, $0\leq \phi \leq \pi$) in the coordinate
system where $\vec{p_3}$ is taken as the $z$ direction and the decay plane 
is identified as $z$-$x$ plane with $ (p_1)_x \geq 0$
(Fig. \ref{fig:coordinate}). 
Note that in this notation the T-odd asymmetry appears as an asymmetry 
in exchange of $x_1$ and $x_2$ 
in the Dalitz plot after integrating out the angle variables.

It is now straightforward to calculate the differential branching ratio
from the effective lagrangian. A detail formula will be given elsewhere
\cite{prep}. There are three types of terms in the differential decay 
branching ratio, {\it i.e.} square of photon-penguin terms,  square of 
four-fermion terms and interference between the 
photon-penguin terms and four-fermion terms.
T-odd terms arise as a part of the interference terms.
These are given by
\begin{eqnarray}
  \label{todd}
  d{\bf B}_{\rm T-odd}
  &=& 
  \frac{3}{16\pi G_F^2}dx_1 dx_2 d\cos\theta d\phi 
        P \sin\theta \sin\phi\
     8 e (x_1 - x_2) \sqrt{\frac{(x_1 + x_2 - 1)}{( 1 - x_1)( 1 - x_2)}}
 \nonumber \\
     &\times&\left[\ 2 (x_1 + x_2 -1)
     {\rm Im}(g_3 A_L^\ast + g_4 A_R^\ast)
     -(2 - x_1 - x_2)
     {\rm Im}(g_5 A_L^\ast + g_6 A_R^\ast)\
     \right],~~~
\end{eqnarray}
where $P$ is the polarization of muons and $e$ is the positron charge.
As seen from the above
expression the T-odd term is generated if the photon-penguin term $A_L$
(or $A_R$) has a different phase from the four-fermion terms $g_3$,
$g_5$ (or $g_4$, $g_6$). These terms are in fact odd 
in exchange of $x_1$ and $x_2$.

In order to extract these T-odd terms from experiments
we also need to know distribution of other terms in the 
$\mu \rightarrow 3e$ differential branching ratio.
In particular, the $|A_L|^2$ and $|A_R|^2$ terms have the following
form. 
\begin{eqnarray}
  \label{photon}
  d {\bf B}_{\rm photon}
  &=&
  \frac{3}{16 \pi G_F^2} dx_1 dx_2 d\cos\theta d\phi
\nonumber \\ 
  &\times& 8 e^2 \left[ \frac{2x_1^2-2x_1+1}{1-x_2} 
                + \frac{2x_2^2-2x_2+1}{1-x_1} \right]
              \left( |A_R|^2 + |A_L|^2 \right),
\end{eqnarray}
where we only keep terms independent of the muon polarization and
neglect the electron mass. The above expression is singular 
for $x_1 \rightarrow 1$ and $x_2 \rightarrow 1$ if we neglect the
electron mass, and therefore the integrated
branching ratio is solely dominated by this kinematical region if
$eA_L$ and $eA_R$ are similar in magnitude as $g_i$'s. In such a case 
the relation ${\bf B}(\mu \rightarrow 3e)/{\bf B}
(\mu \rightarrow e \gamma)\simeq 1/150$ is known to hold \cite{old}.
Since our purpose is to look for the interference between the 
photon-penguin and the four-fermion terms we should exclude
the region very close to $x_1 = 1$ and $x_2 = 1$.
In the actual experiment T-odd asymmetry should be extracted from the
investigation of distribution in the Dalitz plot.
Here in order to present the T-odd effect
we use the branching ratio and asymmetry integrated in the following way.
Defining the regions 
$R_1 = \{x_1 + x_2 \geq 1,\, x_2 \leq 1 - \delta,\, x_2 \geq x_1 \}\ $
and $R_2 = \{x_1 + x_2 \geq 1,\, x_1 \leq 1 - \delta,\, x_1 \geq x_2\}$,
where $\delta$ is introduced to cut off the region $x_1,x_2 \simeq 1$,
the integrated branching ratio and the T-odd asymmetry are defined as
\begin{eqnarray}
  \label{br2}
  {\bf B}[\delta]
 &=&
  \int_{R_1+R_2}dx_1 dx_2 \int^1_{-1} d\cos\theta \int^{\pi}_{0} d\phi
  \frac{d {\bf B}(\mu^+ \rightarrow e^+e^+e^-)}{dx_1 dx_2 d\cos\theta d\phi},
\end{eqnarray}
\begin{eqnarray}
  \label{asymmetry}
  {\bf A}[\delta]
  &=&
  \frac{1}{P\ {\bf B}[\delta]}
  \left[
  \int_{R_1}dx_1 dx_2 \int^1_{-1} d\cos\theta \int^{\pi}_{0} d\phi
  \frac{d {\bf B}(\mu^+ \rightarrow e^+e^+e^-)}{dx_1 dx_2 d\cos\theta d\phi}
  \right.
\nonumber \\
 &-& 
  \left.
  \int_{R_2}dx_1 dx_2 \int^1_{-1} d\cos\theta \int^{\pi}_{0} d\phi
  \frac{d {\bf B}(\mu^+ \rightarrow e^+e^+e^-)}{dx_1 dx_2 d\cos\theta d\phi}
  \right].
\end{eqnarray}

Let us consider the SU(5) SUSY GUT model. In this case the LFV
occurs from the loop effect between the Planck scale
and the GUT scale. Even if we assume that all scalar fields have a
common SUSY breaking mass at the Planck scale, the large top Yukawa
coupling constant becomes a source of the flavor mixing in the 
slepton sector since sleptons belong to the same GUT multiplet 
as squarks \cite{BH,BHS,HMTY2}.
In the simplest SU(5) SUSY GUT the Yukawa 
couplings are given by the following superpotential.
\begin{eqnarray}
  \label{SU5}
  W = T_i (f_u)_{ij} T_j H + T_i (f_d)_{ij} \ovl{F}_j \ovl{H},
\end{eqnarray}
where $T_i$ are $10$ dimensional representations and $\ovl{F}_i$ are the 
$\ovl{5}$ dimensional representations and $H(\ovl{H})$ is $5(\ovl{5})$
dimensional Higgs superfield. Due to the loop effect of $f_u$ the
right-handed stau becomes lighter than other right-handed selectrons at
the GUT scale and therefore slepton mass matrix is no longer simultaneously
diagonalized with lepton mass matrix. In the approximation that the
first two generation's sleptons are degenerate  every term in the LFV
amplitude is proportional to $\lambda_\tau$ $\equiv$ 
$V_R^\ast(e)_{\tau e}V_R(e)_{\tau\mu}$ where $V_R(e)$ is a right-handed 
lepton mixing matrix in the bases where the slepton mass 
matrix is diagonalized up to the left-right mixing terms \cite{BHS}.
Therefore if there are no other source of complex
coupling constants almost no asymmetry is generated
because the photon-penguin and four-fermion terms
have approximately the same phase. Situation will change if we allow complex 
phases for the SUSY breaking terms. Within the assumption of the universal
soft-SUSY breaking terms we can introduce two independent phases which we 
take a phase of the trilinear coupling constant ($A$ term) and a phase of
the higgsino mass term ($\mu$ term).
These phases in general induce large EDMs
of neutron and electron if the masses of squarks or
sleptons are in the range of a few hundred GeV \cite{EDM}. 
In this respect an interesting observation was done in
Ref.\cite{FON} that unlike the $\mu$ phase, the constraint on the $A$
phase is much weaker so that even a $O(1)$ phase is allowed. 
In the followings we assume that the $\mu$ phase vanishes for simplicity 
and see effects of the $A$ phase to the electron and the neutron EDMs and 
the T-odd asymmetry in the $\mu \rightarrow 3e$ decay. 

We calculate the branching ratio Eq.(\ref{br2}) and the T-odd asymmetry
Eq.(\ref{asymmetry}) and the electron and the neutron EDMs in the SUSY 
SU(5) GUT model with a complex $A$ parameter. For calculations of the 
electron and the neutron EDMs in SUSY models we follow Ref.\cite{IN}.
Assuming the universal soft-breaking terms at the Planck scale we solve
the renormalization group equations for the coupling constants and 
SUSY soft-breaking terms from the Planck to the GUT scale, 
and then the GUT to the weak scale. We also require that the electroweak 
symmetry is broken properly due to the renormalization effect and take
into account phenomenological constraints from 
various SUSY particle searches including the constraint from 
the $b \rightarrow s \gamma$ decay as discussed in Ref.\cite{bsll}.
Once we obtain the SUSY breaking terms 
at the weak scale we can evaluate the coupling constants  
$A_R$, $A_L$ and $g_i$'s in Eq.(\ref{lagrangian}) through one loop SUSY
diagrams. $A_R$, $A_L$ are determined by photon-penguin diagrams and
$g_i$'s get contributions from off-shell photon-penguin, $Z$-penguin 
and box diagrams. Formulas for these contributions can be found in
Ref.\cite{HMTYPRD}.

In the calculation of the $\mu \rightarrow e \gamma$ and 
$\mu \rightarrow 3e$ branching ratios there is an important
ambiguity associated with the Yukawa coupling constants at the GUT
scale. As discussed before the branching ratios are 
proportional to $|\lambda_\tau|^2$.
If the Yukawa coupling constants are given only by 
Eq.(\ref{SU5}) the lepton mixing matrix can be 
related to the quark's Cabibbo-Kobayashi-Maskawa (CKM) matrix element. 
It is known, however, that this assumption does not lead to a realistic
mass spectrum for fermions. 
As discussed in Ref.\cite{BH} if other operators are relevant for 
generation of Yukawa coupling constants below the GUT scale the 
relationship between the lepton mixing matrix and the CKM matrix becomes 
quite model-dependent and $\lambda_\tau$ can be
significantly different from the value obtained with the above assumption.
In view of this ambiguity we treat $\lambda_\tau$ as a free parameter and
calculate the $\mu \rightarrow e \gamma$ and $\mu \rightarrow 3e$ 
branching ratios normalized by $|\lambda_\tau|^2$.
Note that the asymmetry and the EDM are essentially independent of 
$\lambda_\tau$.

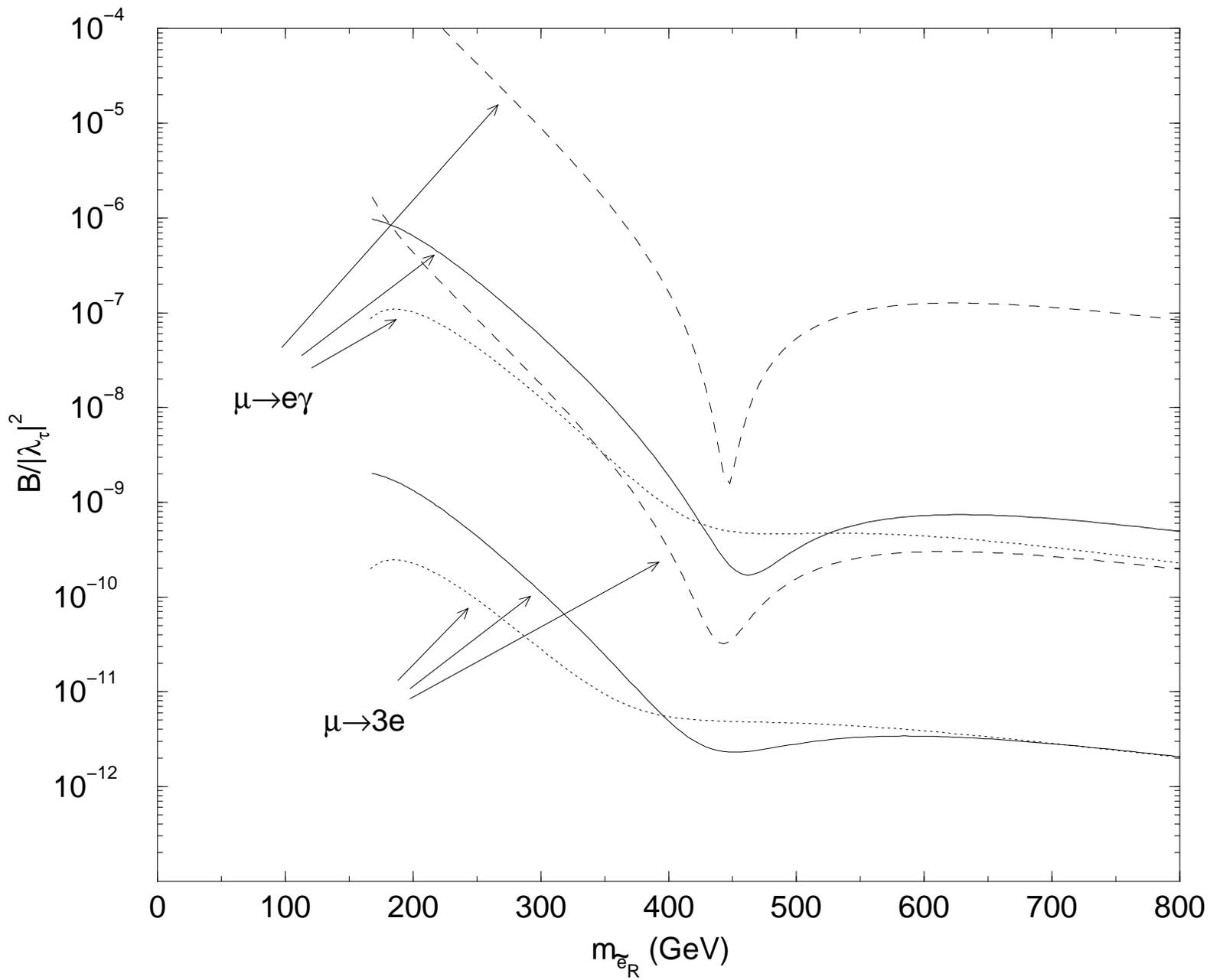
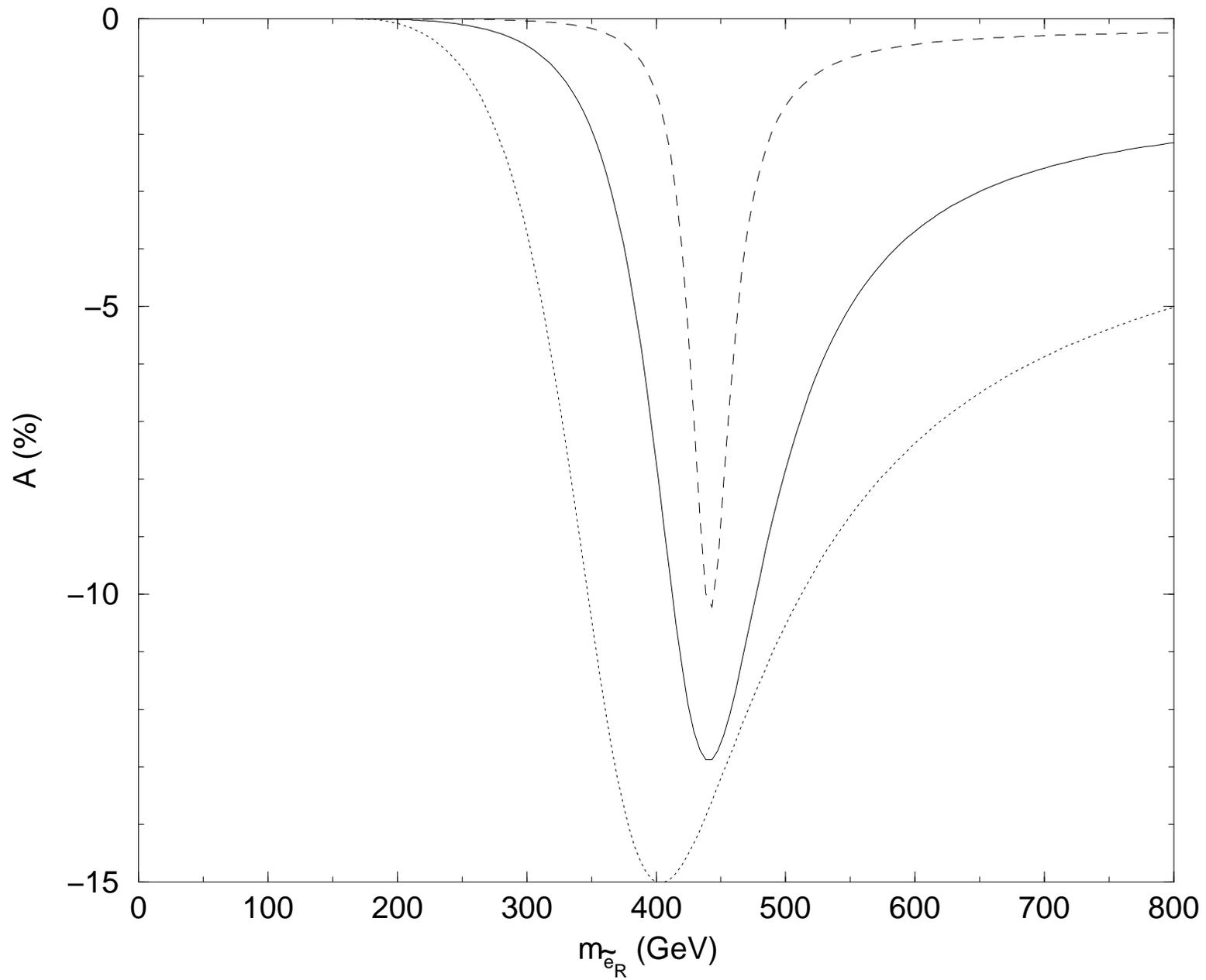
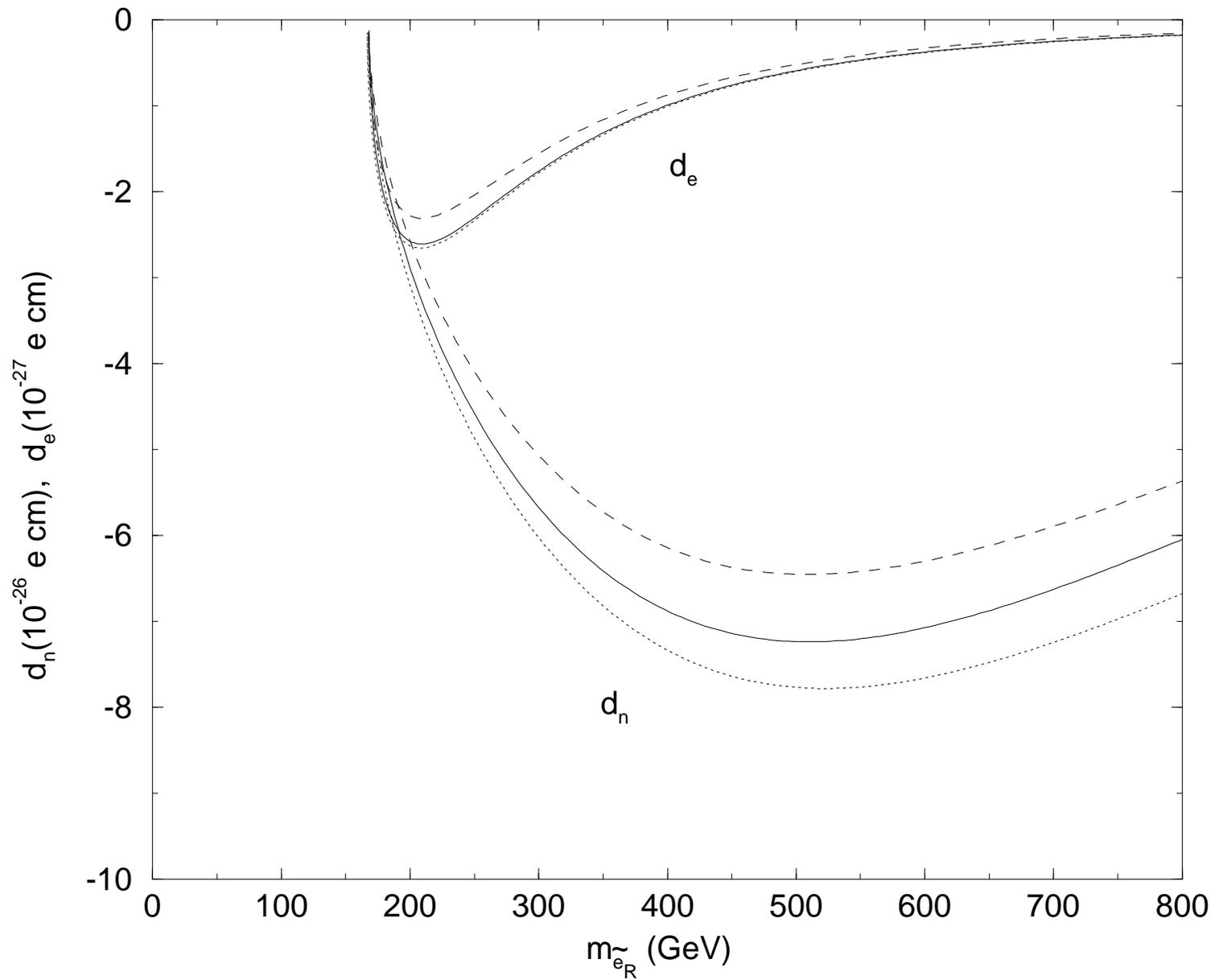
\begin{figure}[htbp]
  \begin{center}
    \leavevmode
    \caption{(a).~Branching ratios for $\mu \rightarrow e \gamma$ and
      $\mu \rightarrow 3e $ normalized by $|\lambda_\tau|^2$$\equiv$ 
      $|V_R^\ast (e)_{\tau e}V_R (e)_{\tau\mu}|^2$ as a function of 
      the right-handed selectron mass $m_{\widetilde e_R}$.
      The cut-off parameter $\delta$ is taken to be 0.02.
      We fix the SUSY parameters as 
      $M_2 = 200$ GeV, $A_X = i$, $\mu > 0$ and $\tan\beta = 3$ (dotted
      line), 10 (solid line), 30 (dashed line) and top quark mass
      as 175 GeV.
      (b).~T-odd asymmetry ${\bf A}$ as a function of 
      $m_{\widetilde e_R}$. Parameters are the same as in (a).
      (c).~The electron and the neutron EDMs as a function of
      $m_{\widetilde e_R}$.}
    \label{fig:lfv}
  \end{center}
\end{figure}

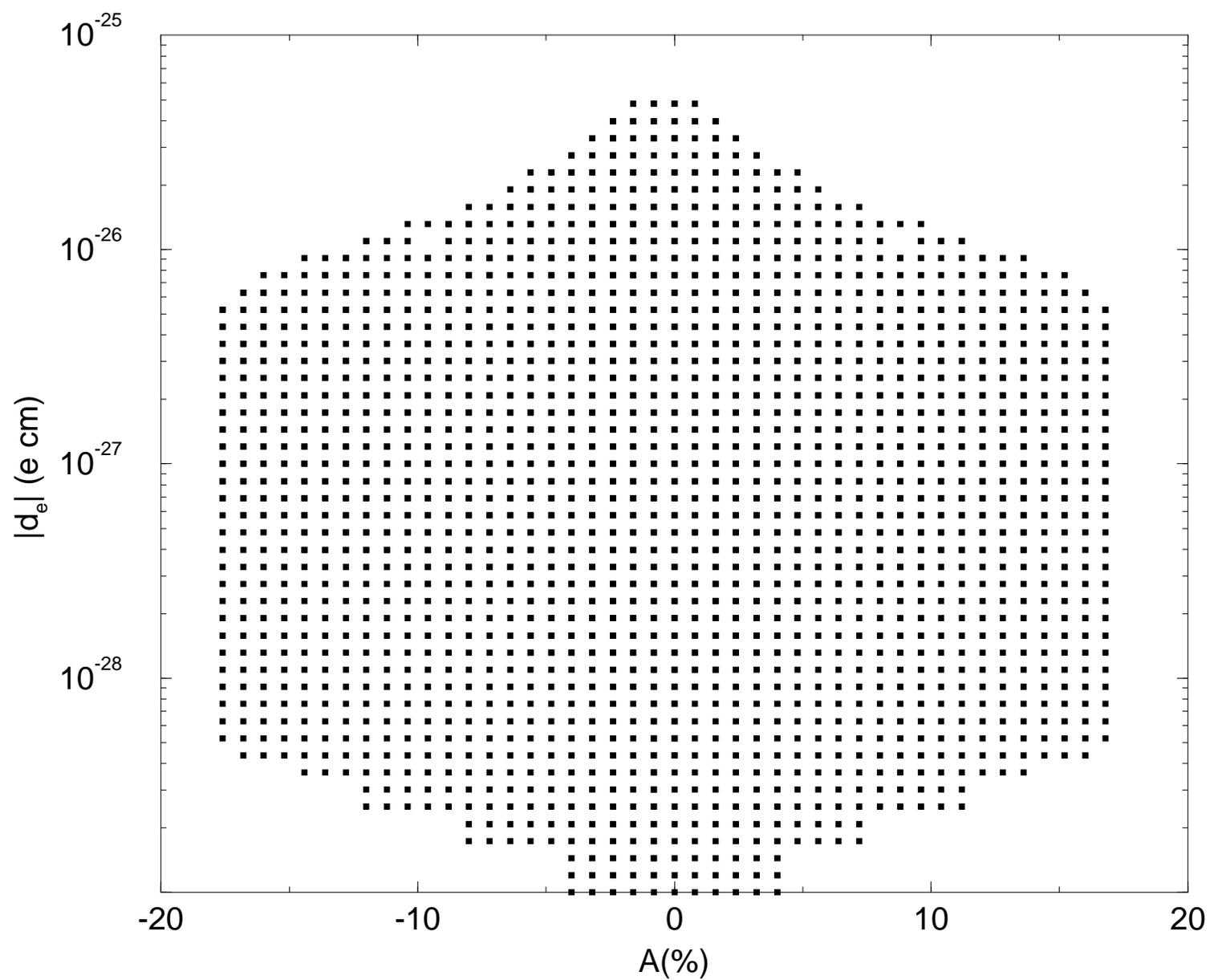
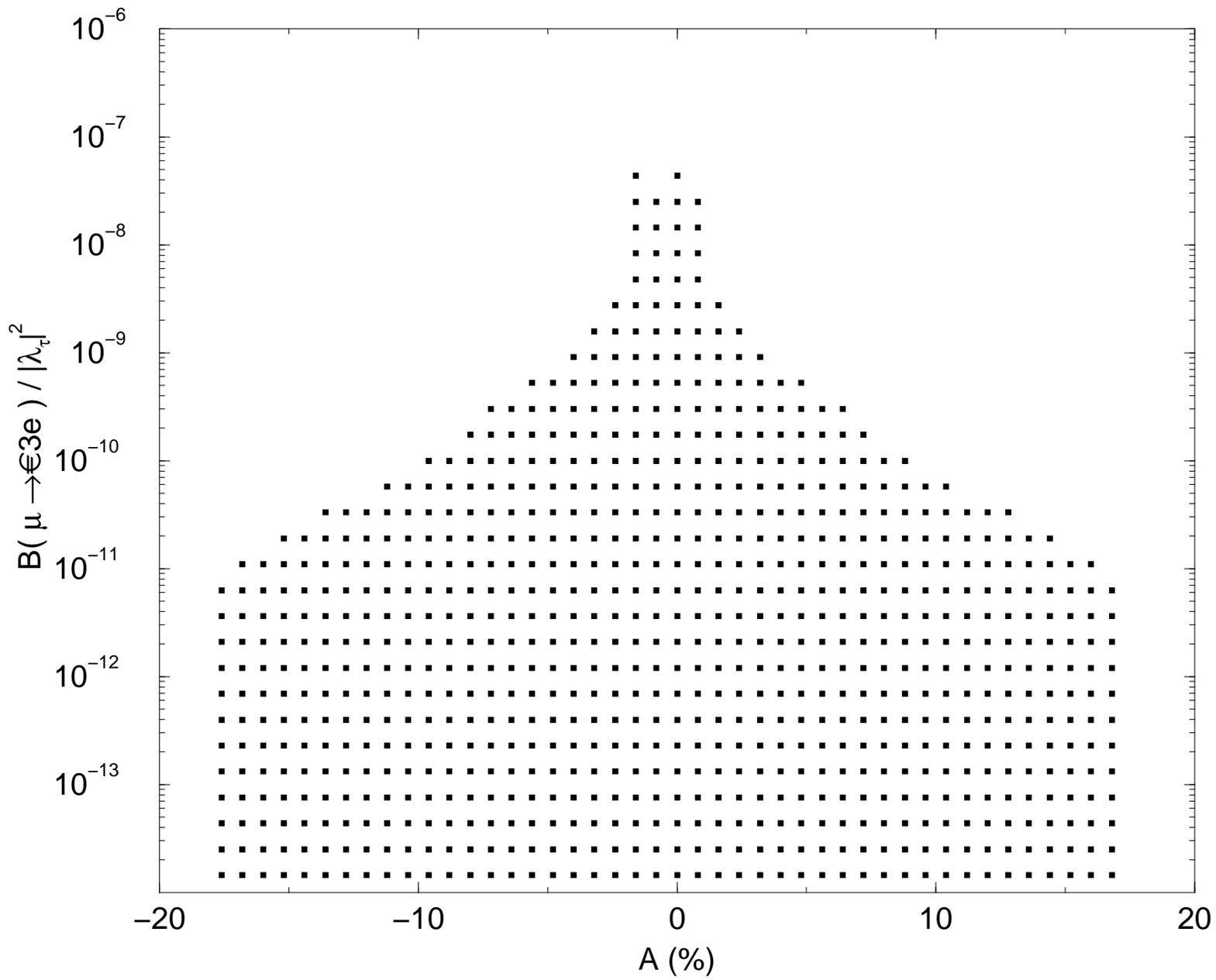
\begin{figure}[htbp]
  \begin{center}
    \leavevmode
    \caption{(a).~A correlation between T-odd asymmetry ${\bf A}$ and the
      absolute value of the electron EDM $|d_e|$.
      The SUSY parameters are scanned in the region $0<m_0<2$ TeV,
      $|A_X|<5$, $0<M_X<2$ TeV and fix arg($A_X)=\pi/2$, $\tan\beta = 5$ and
      $\mu > 0$. 
      (b).~A correlation between T-odd asymmetry ${\bf A}$ and 
      the branching ratio for $\mu \rightarrow 3e $ normalized by 
      $|\lambda_\tau|^2$. The SUSY parameters are the same as in (a) and
      the experimental bounds of the electron and the neutron EDMs are
      imposed.}
    \label{fig:cor}
  \end{center}
\end{figure}

In Fig.\ref{fig:lfv}.(a) we show the branching ratio for 
$\mu \rightarrow e \gamma$ and $\mu \rightarrow 3e$ normalized by
$|\lambda_\tau|^2$ as a function of the right-handed selectron mass 
$(m_{\wt{e}_R})$. 
The SUSY parameters in this model are taken as the SU(2) gaugino mass 
$M_2$, a complex $A$ parameter defined as 
${\cal L}_{soft} = - m_0 A_X(\wt{T}_i(f_u)_{ij}\wt{T}_jH +
\wt{T}_i (f_d)_{ij} \wt{\ovl{F}}_j \ovl{H})$ at the Planck scale, and 
the ratio of two Higgs vacuum expectation values 
($\tan\beta$) and the universal scalar mass $m_0$ at the Planck scale and
the sign of $\mu$. In this figure we fix $M_2=200$ GeV, $A_X=i$ and
$\tan\beta=3,10,30$, $\mu>0$ and the branching ratio are shown as a function 
of $m_{\wt{e}_R}$ instead of $m_0$. The cut off parameter $\delta$
is taken to be $0.02$ for the $\mu \rightarrow 3e$ branching
ratio. We fix the top quark mass as 175 GeV.
If, for example, $|\lambda_\tau| = 1 \times 10^{-2}$ 
the $\mu \rightarrow e\gamma$
branching ratio is $10^{-10}$-$10^{-14}$ and the $\mu \rightarrow 3e$
branching ratio is $10^{-12}$-$10^{-16}$ for $\tan\beta=10$.
On the other hand if $\lambda_\tau$ is given by the corresponding CKM matrix 
elements $V_{td}^\ast V_{ts}$ then 
$|\lambda_\tau| = (3-5)\times 10^{-4}$ and
therefore the branching ratio is smaller by three orders of magnitude.
In Fig.\ref{fig:lfv}.(b) the T-odd asymmetry defined in 
Eq.(\ref{asymmetry}) is shown for the same parameters as in 
Fig.\ref{fig:lfv}.(a). Also the electron and the neutron EDMs are shown
in Fig.\ref{fig:lfv}.(c).
We can see that the asymmetry becomes maximal around 
$m_{\wt{e}_R} = 400$ GeV while the magnitude of the electron (neutron)
EDM is a decreasing function of $m_{\wt{e}_R}$ above 200 (400) GeV.
The difference of these behaviors comes from the fact that for the asymmetry 
photon-penguin and four-fermion terms have similar magnitude for
$m_{\wt{e}_R} \simeq 400$ GeV 
but for the EDMs only photon-penguin diagram is relevant.
It is interesting to see that when the asymmetry is large the ratio of the 
${\bf B}(\mu \rightarrow e \gamma)$ and ${\bf B}(\mu \rightarrow 3e)$
is $1/50\ $-$\ 1/100$.
Fig.\ref{fig:cor}.(a) shows the correlation between the absolute value of
the electron EDM and the T-odd asymmetry. 
In this figure, the SUSY parameters are scanned in the
region $0< m_0 < 2$ TeV, $|A_X|<5$, $0 < M_X < 2$ TeV and fix arg($A_X$)
=$\pi/2$, $\mu>0$.
We find that the sign of the EDM and that of
asymmetry are the same but here the absolute value of the EDM is shown.
We can see that the electron EDM and the T-odd asymmetry do not show
strong correlation. 
Within the current experimental upper bound of the EDM, which is
$|d_e| < 4 \times 10^{-27}$ \cite{de}, 
the asymmetry can be as large as 18 \%.
This means that the T-odd asymmetry in the 
$\mu \rightarrow 3e$ process gives independent information from the
electron EDM on the possible complex parameters in the SUSY  GUT model.
Correlation between the neutron EDM and T-odd asymmetry is also
calculated. As in Fig.\ref{fig:cor}.(a) we do not see any correlation and
the $\pm 18 \%$ asymmetry is possible within the experimental upper
bound of the neutron EDM ($|d_n|<1.1 \times 10^{-25}$ \cite{dn}).
Fig.\ref{fig:cor}.(b) shows that the correlation between 
${\bf B}(\mu \to 3e)/|\lambda_\tau|^2$ and T-odd asymmetry.
In this figure we take into account the experimental bounds of the
electron and the neutron EDMs. We see that, for example, when
$|\lambda|$ is $1 \times 10^{-2}$ $10\%$ asymmetry is possible for 
${\bf B}(\mu \to 3e) \simeq 10^{-14}$.
We also investigated the correlation between the asymmetry and the
slepton mass and found that the slepton mass which
corresponds to maximal asymmetry changes from 200 GeV to 2 TeV
when we scanned the parameters in the above region,
so that a large asymmetry does not necessarily mean a light slepton.

Let us discuss T-odd asymmetry in SUSY models 
other than SU(5) GUT. In the minimal SO(10) SUSY GUT model 
it is pointed out in Ref.\cite{BHS} that the $\mu \rightarrow 
e \gamma$ branching ratio is enhanced by $(m_\tau/m_\mu)^2$ compared 
to the SU(5) model. The $\mu \rightarrow 3 e$
branching ratio is also enhanced by the same amount and in this
case the photon-penguin and four-fermion terms can have different phases
without the phases of the soft breaking terms.
But unfortunately, the T-odd asymmetry cannot be large 
because only the photon-penguin term is enhanced in this decay. 
Another possibility to induce large LFV is the SUSY model with 
the right-handed neutrino supermultiplet \cite{HMTYPRD,HMTY}.
In this case the right-handed neutrino's Yukawa coupling constants 
become new sources of LFV and CP violation.
This model is also interesting in connection with the baryon number 
asymmetry of the Universe since the right-handed neutrino or
right-handed sneutrino decays can be the origin of the leptogenesis
\cite{leptogenesis} and CP violation in the Yukawa coupling constants is one of
the necessary ingredients.
We have investigated the LFV branching ratio 
and the T-odd asymmetry in this model.
We take neutrino masses in the range of 10$^{-3}$ -
10 eV and adjust the right-handed neutrino Majorana mass to $10^{12}$ -
$10^{16}$ GeV to get $O(1)$  Yukawa coupling constants. 
In this model using the freedom of right-handed Yukawa coupling
constants, it is possible to generate two independent phases
in the photon-penguin terms and the four-fermion terms without help of 
the complex soft breaking terms.
With real soft-breaking terms
typical magnitude of asymmetry turns out to be less than 0.1 \%
although it is possible to have an asymmetry up to 10 \% by tuning the
parameters in the right-handed Yukawa couplings.

We only consider the polarized muon decay here. Extension to
the tau decay is straightforward. For example the T-odd asymmetry 
in $\tau \rightarrow 3 \mu$ is obtained by replacing relevant generation 
indices in the  $\mu \rightarrow 3 e$ formula. It should be noticed,
however, that there is essentially no difference between SU(5)
and SO(10) models in this case because there are no enhancement mechanism
for the SO(10) model compared to the SU(5) case in this decay mode.
 
In this letter we investigate a possibility to observe a sizable CP
nonconserving asymmetry in the $\mu \rightarrow 3e$ decay in SUSY models. 
In the SU(5) GUT model sizable LFV arises from the GUT  interactions and 
an asymmetry up to 18 \% can be induced by the complex phase of
trilinear soft-breaking term.
In order to measure the asymmetry of this magnitude
the  the $\mu \rightarrow 3e$ branching ratio has to be large enough.
In the SU(5) model the magnitude of the branching ratio itself strongly
depends on the model-dependent parameter $\lambda_\tau$ so that
practically the  $\mu \rightarrow 3e$ branching ratio is only
constrained by the experimental bounds of the $\mu \rightarrow 3e$ 
and $\mu \rightarrow e\gamma$ processes.
Therefore if the $\mu \rightarrow e\gamma$ process is observed at the
level of $10^{-12}$-$10^{-11}$ in near future, 
the $\mu \rightarrow 3e$ branching ratio can be $O(10^{-13})$
and an experiment of $\mu \rightarrow 3e$ with a sensitivity of 
the $10^{-15}$ level could reveal the T-odd asymmetry.
Note that stronger bounds on EDM or slepton mass do not mean that
the asymmetry cannot be measurable because due to ambiguity of
$\lambda_\tau$ the branching ratio cannot be strongly constrained by 
these bounds.
Besides the EDMs for neutron and electron, the T-odd asymmetry therefore
gives us a new possibility to search for CP violation in SUSY models.

The authors would like to thank Y.~Kuno and K.~Tobe for  useful comments.
This work was supported in part by the
Grant-in-Aid of the Ministry of Education, Science, Sports and
Culture, Government of Japan.


\newpage

\hspace{-1.2cm}{\Large {\bf Figure Captions:}}

\end{document}